\documentclass[preprint,showpacs,preprintnumbers,amsmath,amssymb]{revtex4}

\usepackage{graphicx}
\usepackage{dcolumn}
\usepackage{bm}

\begin{document}


\title{Optimal Colored Noise for Estimating Phase Response Curves}

\author{Kazuhiko Morinaga}
\author{Ryota Miyata}%
\author{Toru Aonishi}
\email{aonishi@dis.titech.ac.jp}
\affiliation{Interdisciplinary Graduate School of Science and Engineering,
Tokyo Institute of Technology, Yokohama, 226--8502, Japan
}%

\date{\today}

\begin{abstract}
The phase response curve (PRC) is an important measure representing the interaction between oscillatory elements. To understand synchrony in biological systems, many research groups have sought to measure PRCs directly from biological cells including neurons. Ermentrout et al. and Ota et al. showed that PRCs can be identified through measurement of white-noise spike-triggered averages. The disadvantage of this method is that one has to collect more than ten-thousand spikes to ensure the accuracy of the estimate. In this paper, to achieve a more accurate estimation of PRCs with a limited sample size, we use colored noise, which has recently drawn attention because of its unique effect on dynamical systems. We numerically show that there is an optimal colored noise to estimate PRCs in the most rigorous fashion.
\end{abstract}

\pacs{05.45.Xt, 87.18.Tt, 87.19.lm}
\maketitle

\section{Introduction}

Understanding oscillatory phenomena is one of key issues in various research fields \cite{pikovsky,buzsaki}. According to the theory of phase reduction, interactions between elements of a large-scale oscillatory system are often formalized in terms of phase response curves (PRCs) \cite{kuramoto}. Recently, many research groups have sought to measure PRCs directly from real systems, especially biological oscillators such as circadian clocks \cite{ukai} and neural oscillators \cite{reyes,schultheiss}. However, it is difficult to perform perturbation experiments on such biological systems because of the non-stationarity of the oscillation and the shortness of the survival time. Many methods have been proposed to overcome the experimental limitations \cite{nielsen,schultheiss}. In particular, Ermentrout et al. analytically derived a relational expression between the PRC and the white-noise spike-triggered average (wSTA) \cite{ermentrout1}. The STA defined as the average stimulus preceding a spike provides an estimate of stimulus features encoded by neurons. Their study clarified the relationship between neural dynamics and neural coding and simultaneously showed the possibility of identifying PRCs by measuring wSTAs. Ota et al. extended this work and showed the effectiveness of the PRC estimation via the wSTA measurement \cite{kota1}. The advantages of this method are that 1) the experiment can be realized with an open loop system, and thus it does not require special equipment like the dynamic clamp \cite{dorval,ota2}, and that 2) this method can be applied to non-stationary situations such as switching between regular and burst-firing modes. On the other hand, its disadvantage is that one has to collect many samples (more than ten-thousand spikes) to ensure the accuracy of the estimate, as Fig. 1 of \cite{ermentrout1} and below show. 

Recently, various nontrivial phenomena caused by colored noise have started to be investigated in the field of dynamical systems \cite{nakao,simakov,kurebayashi}. In this paper, we use colored noise to achieve a more effective estimate of PRCs. We derive an equation that relates the PRC and colored-noise spike-triggered average (cSTA) and propose a simple statistical method for estimating PRCs from cSTAs on the basis of this relation. Note that Ermentrout et al. derived a general relational expression that includes the colored case, but their theoretical results are different from ours. We numerically demonstrate that we can ensure a more accurate PRC estimate when cSTAs are measured with a smaller sample size than that of a wSTA measurement and that there is an optimal colored noise to estimate PRCs in the most rigorous fashion for a limited sample size.

\section{Colored-noise spike-triggered average (cSTA)}

In acquiring STAs in white and colored noise cases, we measure the spike time $t_s$ of a neuron while storing the noise stimulus $\sigma \xi(t)$. In the following derivation, we assume that the noise stimulus can be scaled by $\sigma$ in order to clarify the order of each expansion term. Under this postulate, the STA $S(t)$ is defined as the ensemble average of the stimuli preceding the spike time $t_s$: $S(t)=\sigma \left<\xi(t_{s}-t)\right>$ \cite{ermentrout1,schwartz,ota1}. Denoting the time of the $i$th spike as $t_{s}^i$ and the time sequence of the noise stimulus preceding the $i$th spike as $\sigma \xi(t_{s}^i-t)$, $(t>0)$, we can estimate the STA by taking the empirical mean of the stimuli \cite{schwartz,ota1}:
\begin{eqnarray}
\hat{S}(t) = \frac{\sigma}{N} \sum_{i=1}^N \xi\left(t_{s}^i-t\right), \label{eq.0}
\end{eqnarray}
where $N$ indicates the number of samples. In this paper, $\hat{S}(t)$ is called empirical STA. 

We will focus on the case of a limit-cycle oscillator including a neuron with its own natural period of $T$ stimulated by noise $\sigma \xi(t)$. If the amplitude of the noise, $\sigma$, is sufficiently small, one can describe the evolution of the oscillator perturbed by it as follows \cite{kuramoto,ota1,ermentrout1,kota1}:
\begin{eqnarray}
\frac{d\phi}{dt} = 1 + \sigma Z(\phi) \xi(t),
\label{eq.1}
\end{eqnarray}
where $\phi$ is the phase, $\xi(t)$ represents the noise stimulus described above, and $Z(\phi)$ is the phase response curve. For mathematical and numerical tractability, the noise stimulus $\xi(t)$ is assumed to be generated with a simple Ornstein-Uhlenbeck (OU) process \cite{risken}:
\begin{eqnarray}
\tau \frac{d\xi}{dt} = - \xi + \Gamma(t), 
\label{eq.2} 
\end{eqnarray}
where $\Gamma(t)$ is white noise satisfying $\left<\Gamma(t)\right>=0$ and $\left<\Gamma(t)\Gamma(t')\right>= \delta(t-t')$. The above OU process can generate a colored noise signal, whose auto-correlation function $K(t)$ and power spectrum $P(\omega)$ are
\begin{eqnarray}
K(t) &=& \frac{1}{2\tau} \exp\left(-\frac{|t|}{\tau} \right), \label{eq.3} \\
P(\omega) &=& \frac{\tau}{1 + \omega^2 \tau^2}, \label{eq.4}
\end{eqnarray}
where $\omega$ is angular frequency. Here, $\left<\xi(t)\right> = 0$. The colored noise used here is categorized as red noise in which the corner angular frequency of its spectrum is $\tau$. Furthermore, the total power of this noise, which is defined as $\int_{-\infty}^{+\infty} P(\omega)d\omega$, is invariant under a change of $\tau$. Note that the theory derived below can be applied to other noise models if their auto-correlation functions decay faster than $1/t$. 

Hereinafter, we analytically derive cSTA in a different way from Ermentrout's derivation of STA. We deal with long time correlations of noise. First, we introduce a slow phase variable $\psi(t)$ as $\phi(t) = t + \psi(t)$, as well as a spike time $t_s$ satisfying the termination condition $\phi(t_s) = t_s + \psi(t_s) = nT$. $n$ is a positive large integer at which the OU process is in equilibrium. The slow phase $\psi$ at $t_s - t'$ before $t_s$ obeys the following reverse-time phase equation:
\begin{eqnarray}
- \frac{d\psi(t_s-t')}{dt'} = \sigma Z(t_s - t' + \psi(t_s-t')) \xi(t_s-t'). \label{eq.5}
\end{eqnarray}
The above reverse-time phase equation can be solved by integration with respect to $t'$ from $0$ to $t_c$ as follows:
\begin{eqnarray}
\psi(t_s) - \psi(t_s -t_c) = \sigma \int_{0}^{t_c} dt' Z(t_s-t'+\psi(t_s-t'))\xi(t_s-t'). \label{eq.6}
\end{eqnarray}
Here, we can select $t_c$ satisfying $\psi(t_s -t_c)=0$ without loss of generality on the condition that $t_c$ is sufficiently larger than $\tau$ such that $K(t_c) \sim 0$. This is because if $t_c>>\tau$, the system has lost memory of the termination condition at $t=t_s$. Since $\psi(t_s-t')$ varies slowly when $\sigma<<1$, we can expand the right-hand side of Eq. (\ref{eq.6}) by following the recipe in \cite{ota1} and obtain the lowest order term scaled by $\sigma$ as follows:
\begin{eqnarray}
\psi(t_s) &=& \sigma \int_{0}^{t_c} dt' Z(t_s-t'+\psi(t_s-t'))\xi(t_s-t') \nonumber \\
&=& \sigma \int_{0}^{t_c} dt' Z(t_s-t'+\psi(t_s))\xi(t_s-t') + O(\sigma^2) \nonumber \\
&=& \sigma \int_{0}^{t_c} dt' Z(t_s-t')\xi(t_s-t') + O(\sigma^2).
 \label{eq.7}
\end{eqnarray}
The last equation above can be derived from the middle one because $\psi(t_s)$ is $O(\sigma)$ if $\psi(t_s -t_c)=0$. 

On the basis of the formulation in \cite{ota1}, STA including both white and colored cases can be described by
\begin{eqnarray}
S(t)  &=& \sigma \int_{-\infty}^{\infty} dt_s \left<\delta\left(t_s + \psi(t_s) - nT\right) \xi(t_s - t)\right>. \label{eq.8}
\end{eqnarray}
Here, $\delta\left(t_s + \psi(t_s) - nT\right)$ represents the restraint condition of the spike time $t_s$. Equation (\ref{eq.8}) means that only samples of $\xi(t_s-t)$ satisfying the termination condition, $t_s + \psi(t_s) = nT$, are averaged. Next, substituting Eq. (\ref{eq.7}) into Eq. (\ref{eq.8}), we expand Eq. (\ref{eq.8}) as follows: 
\begin{eqnarray}
& &S(t) = \sigma \int_{-\infty}^{\infty} dt_s \delta\left(t_s - nT\right) \left<\xi(t_s - t)\right> \nonumber \\
& &+\sigma^2 \int_{-\infty}^{\infty} dt_s \delta'\left(t_s - nT\right) \int_{0}^{t_c} dt' Z(t_s-t')\left<\xi(t_s-t')\xi(t_s - t)\right> + O(\sigma^3), \label{eq.9}
\end{eqnarray}
where $\left<\xi(t_s - t)\right>=0$ and $\left<\xi(t_s-t')\xi(t_s - t)\right>=K(t-t')$ due to the definition of colored noise. Finally, taking the integration by parts in the second term of Eq. (\ref{eq.9}), we obtain the main result:
\begin{eqnarray}
S(t) = - \sigma^2 \int_{0}^{\infty} dt' Z'(T-t')K(t-t'). \label{eq.10}
\end{eqnarray}
Here, $Z'(t)$ is a periodic function satisfying $Z'(nT-t)=Z'(T-t)$, and $t_c$ can be safely replaced with infinity because $t_c$ is sufficiently larger than $\tau$ such that $K(t_c)\sim 0$. 

\begin{figure}
\includegraphics[width=15.0cm]{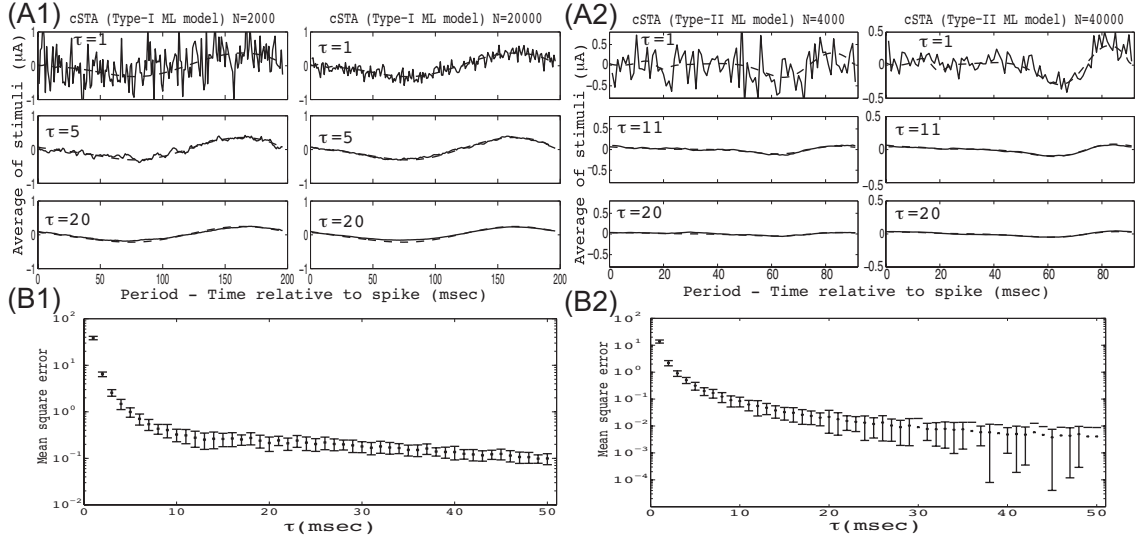}
\caption{\label{fig1}
The stochastic convergence of empirical cSTA to the theoretically derived one becomes faster as the time constant of the noise stimulus, $\tau$, increases, whereas the total power of the noise stimulus is invariant. (A1) (A2) Snapshots of empirical cSTAs (solid lines) and theoretically derived ones (dashed lines) for different values of $\tau$. (A1) Type-I ML model. Left row: $N = 2000$. Right row: $N = 20000$. (A2) Type-II ML model. Left row: $N = 4000$. Right row: $N = 40000$. (B1) (B2) The mean square error (MSE) between the empirical cSTA and the theoretical one as a function of $\tau$. Dotted points and error bars respectively show the averages and standard deviations of MSE. (B1) Type-I ML model. 50 simulations. $N=2000$. (B2) Type-II ML model. 100 simulations. $N=4000$. In the Type-I ML model, $T = 195.83$ [msec] and $\sigma=1$ [$\mu A$]. In the Type-II ML model, $T = 91.17$ [msec] and $\sigma = \sqrt{1.5}$ [$\mu A$].
}
\end{figure}

To verify our theory, we compared theoretically derived cSTAs with empirical cSTAs calculated from finite samples generated by numerical simulations. To simulate experiments with colored noise stimuli, we used Euler's method to solve Type-I and Type-II Morris-Lecar (ML) models \cite{morris,ermentrout2} and the OU process in Eq. (\ref{eq.2}). Moreover, to calculate theoretical cSTAs, we numerically obtained highly accurate PRCs of the Type-I and Type-II ML models by using the direct method \cite{schultheiss}. Figures \ref{fig1} (A1) and (A2) shows the theoretically predicted (using Eq. (\ref{eq.10})) and empirical cSTAs for different noise time constants and sample sizes. The theoretical cSTAs are in good agreement with the empirical one. Furthermore, the results shown in Figs. \ref{fig1}(A1) and (A2) suggest that the amplitude of fluctuations of empirical cSTA might depend on the noise time constant $\tau$. Next, we evaluated the degree of convergence of empirical cSTA for various noise time constants. Note that as described above, the total power of this noise is invariant under a change of the time constant. Here, to evaluate convergence, we measured the mean square error (MSE) between the theoretical cSTA and the empirical cSTA. As shown in Figs. \ref{fig1}(B1) and (B2), for the same number of samples, the MSE decreases as the noise time constant increases. These results suggest that the stochastic convergence of empirical cSTA becomes faster as the time constant increases, whereas the total power of the noise stimulus is invariant. Note that we confirmed that the results shown here are consistent with those obtained from other models such as the Connor-Stevens (CS) model \cite{ermentrout2}. 

\begin{figure}
\includegraphics[width=7.0cm]{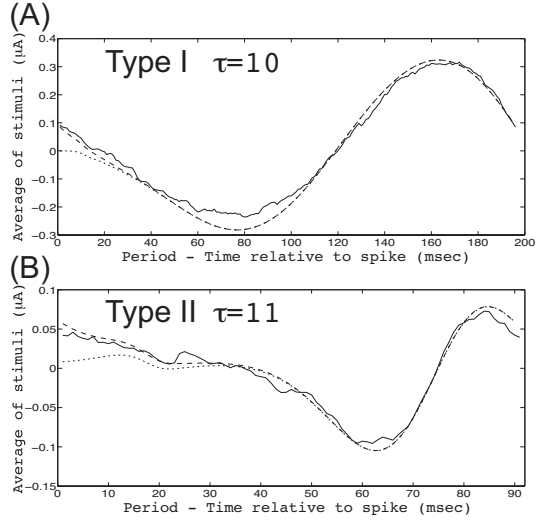}
\caption{\label{fig2} 
Difference between theoretical cSTAs obtained from Eq. (\ref{eq.10}) in this paper and Eq. (4) in \cite{ermentrout1}. (A) (B) Snapshots of empirical cSTAs (solid lines) and theoretical cSTAs calculated with Eq. (\ref{eq.10}) in this paper (dashed lines) and Eq. (4) in \cite{ermentrout1} (dotted lines). (A) Type-I ML model. $\tau=10$ [msec]. $N=20000$. $T = 195.83$ [msec]. $\sigma=1$ [$\mu$ A]. (B) Type-II ML model. $\tau=11$ [msec]. $N=40000$. $T = 91.17$ [msec]. $\sigma = \sqrt{1.5}$ [$\mu$ A].
}
\end{figure}

Ermentrout et al. derived the theoretical STA of oscillators, including the colored noise case \cite{ermentrout1}. In the limit of $\tau \rightarrow 0$ (i.e. white noise), our main result, Eq.  (\ref{eq.10}), becomes $S(t)=-\sigma^2 Z'(T-t)$ and is identical to Eq. (4) in \cite{ermentrout1}. However, when $\tau$ is finite, our result is different from theirs. To prove there is a difference between Eq. (\ref{eq.10}) in this paper and Eq. (4) in \cite{ermentrout1}, we draw graphs consisting of two different theoretical cSTAs obtained from these two equations and empirical cSTAs in the Type I and II ML models (Fig. \ref{fig2}). As shown in Fig. \ref{fig2}, our theoretical cSTAs are in better agreement with the empirical ones than those obtained by using Eq. (4) in \cite{ermentrout1}. In particular, their derivation assumed that the input within the period between two successive spikes could only affect the latter of the two spikes, and thus, it does not deal with long-duration correlations of the noise.

\begin{figure}
\includegraphics[width=15.0cm]{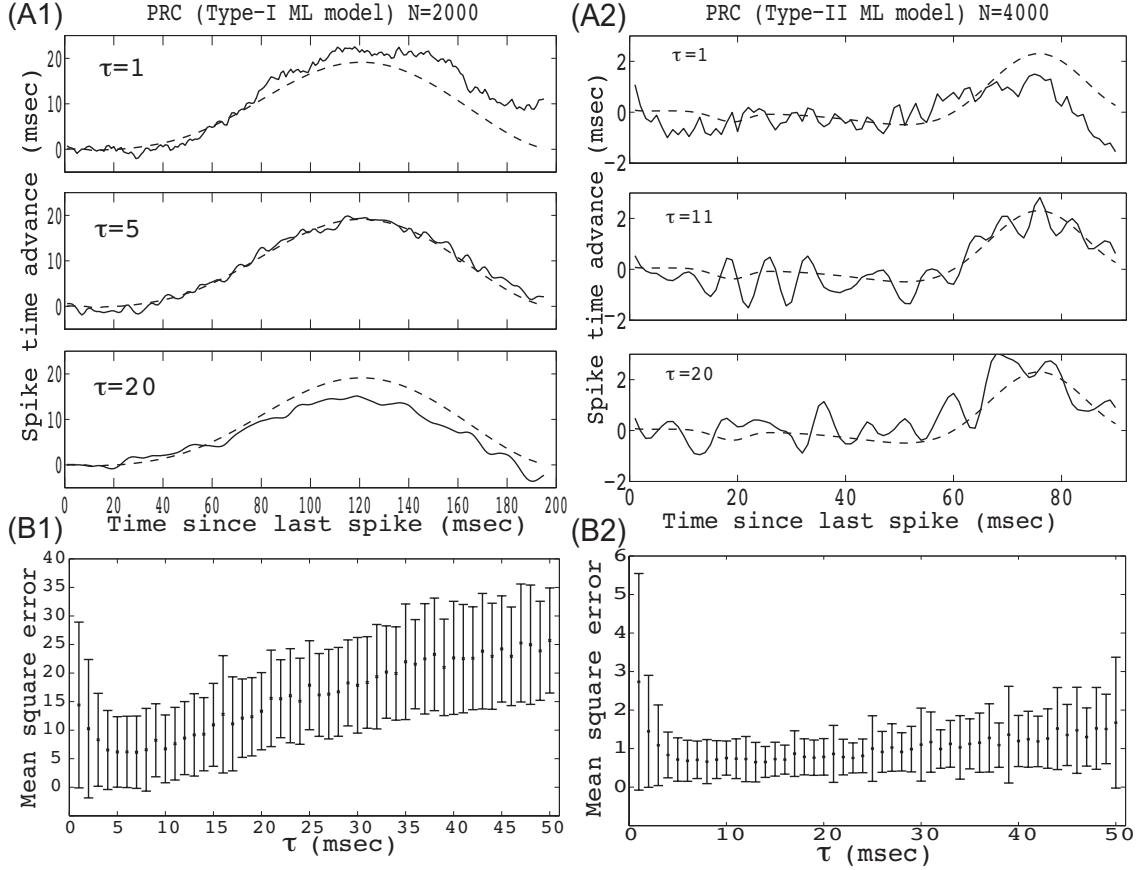}
\caption{\label{fig3} There is an optimal value of the noise time constant to obtain the best estimate of PRC. (A1) (A2) Least squared estimation of PRCs from empirical cSTAs with various $\tau$. Snapshots of estimated PRCs (solid lines) from empirical cSTAs and PRCs (dashed lines) obtained with the direct method. (A1) Type-I ML model. $N=2000$. (A2) Type-II ML model. $N=4000$. (B1) (B2) MSE between the PRC obtained with the direct method and the estimated one as a function of $\tau$. Dotted points and error bars respectively show the averages and standard deviations of MSE. (B1) Type-I ML model. 100 simulations. $N=2000$. (B2) Type-II ML model. 100 simulations. $N=4000$. In the Type-I ML model, $T = 195.83$ [msec] and $\sigma=1$ [$\mu A$]. In the Type-II ML model, $T = 91.17$ [msec] and $\sigma = \sqrt{1.5}$ [$\mu A$].}
\end{figure}

\section{ Estimation of PRC from empirical cSTA}

Next, we will construct a simple algorithm to estimate PRCs from empirical cSTAs. In particular, we construct a least-squares algorithm \cite{bishop} on the basis of Eq. (\ref{eq.10}). We discretize Eq. (\ref{eq.10}) in time and approximate the integral in this equation using the rectangle method: 
\begin{eqnarray}
S_d(nh) =- \sigma^2 h \sum_{m=0}^{L-1}Z'(T-mh) \tilde{K} \left((n-m)h\right), \label{eq.11}
\end{eqnarray} 
where $h = L/T$ and $\tilde{K}(t)= \sum_{j=0}^{\infty}K\left(t-jT\right)$. Here, the interval of integration changes from $[0,+\infty)$ to $[0,T]$ by introducing $\tilde{K}(t)$. In the limit of $h\rightarrow 0$, Eq. (\ref{eq.11}) is exactly equal to Eq. (\ref{eq.10}). Accordingly, the square error between a spike-triggered stimulus $\sigma \xi(t_{s}^{i}-nh)$ and cSTA $S_d(nh)$ can be written as
\begin{eqnarray}
J = \sum_{i=1}^{N} \sum_{n=0}^{L-1} \left(\sigma \xi\left(t_{s}^{i}-nh\right) - S_d(nh)\right)^2. \label{eq.12}
\end{eqnarray}
The point of minimum error satisfies the following simultaneous linear equation,
\begin{eqnarray}
\left[ 
\begin{array}{c}
B_{0} \\
 \vdots\\
B_{L-1}\\
\end{array} 
\right]
= - \sigma h
\left[ 
\begin{array}{ccc}
C_{0,0}&\cdots &C_{0,L-1} \\
\vdots&\ddots & \vdots\\
C_{L-1,0}&\cdots&C_{L-1,L-1} \\
\end{array} 
\right]
\left[ 
\begin{array}{c}
\hat{Z}'(h) \\
 \vdots\\
\hat{Z}'(Lh)\\
\end{array} 
\right],
\label{eq.13}
\end{eqnarray}
where $B_k=\sum_{n=0}^{L-1}\hat{S}\left(nh\right)\tilde{K}\left((n-k)h\right)$ and $C_{m,k} = \sum_{n=0}^{L-1}\tilde{K}\left((n-m)h\right)\tilde{K}\left((n-k)h\right)$. Here, $\hat{S}\left(nh\right)$ is the empirical cSTA defined in Eq. (\ref{eq.0}), and the matrix $[C_{m,k}]$ is a positive symmetric matrix. Thus, by plugging the empirical cSTA in Eq. (\ref{eq.13}), we can find a unique solution $\hat{Z}'(kh)$ $(k=1,\cdots,L)$. Note that the computational time of this algorithm is independent of the sample size $N$, because $B_k$ is calculated as the convolution between empirical cSTA already calculated and $\tilde K$. Finally, integrating the estimate $\hat{Z}'(kh)$ numerically, we obtain an estimated PRC $\hat{Z}(kh)$.

We compared PRCs and estimates made with this algorithm from empirical cSTAs. As in the above numerical experiment, we numerically obtained highly accurate PRCs of the Type-I and Type-II ML models by using the direct method. Figures \ref{fig3}(A1) and (A2) show PRCs obtained with the direct method and PRCs estimated from empirical cSTAs for different noise time constants. The estimates at $\tau=5$[msec] and $\tau=11$[msec] fit the PRC obtained with the direct method the best. Thus, the snapshots of the estimates shown in Figs. \ref{fig3} (A1) and (A2) suggest that we can accurately estimate the PRC if we select an appropriate value of the noise time constant. Next, we evaluated the accuracy of the estimated PRC for various noise time constants. Here, to evaluate accuracy, we measured the MSE between the PRC $Z(t)$ obtained with the direct method and the estimate PRC $\hat Z(t)$. Figures \ref{fig3} (B1) and (B2) plot MSEs as a function of the time constant. The results suggest that the PRC can be estimated more accurately by using colored noise than by using white noise. Furthermore, as can be seen from Figs. \ref{fig3} (B1) and (B2), there is an optimal value for the noise time constant to obtain the best estimate of PRC. Note that we confirmed that the results shown here are consistent with those obtained from other models such as the CS model.

\section{Discussion}

\begin{figure}
\includegraphics[width=7.0cm]{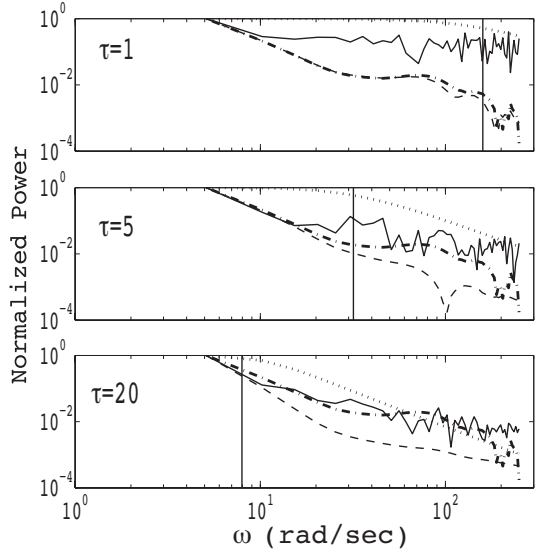}
\caption{\label{fig4} The high-frequency components of noise stimuli mainly contribute to the error of the empirical cSTA. Normalized power spectra of empirical cSTAs (solid lines), theoretical cSTAs (dashed lines), and noise stimuli (bold dotted lines) for different values of $\tau$. For comparison, the normalized power spectrum of theoretical wSTA (bold dashed-dotted line) is superimposed on each of these graphs. Type-I ML model. $T = 195.83$ [msec]. $\sigma=1$ [$\mu A$]. $N=2000$.}
\end{figure}

Figure \ref{fig4} shows the normalized power spectra of empirical and theoretical cSTAs, noise stimuli, and theoretical wSTA in the Type-I ML model. As can be seen, the higher frequency components of empirical cSTA deviate from the theoretical ones when $N=2000$. Note that the error of these components decreases in proportion to $1/N$ for any value of $\tau$ (data not shown). This result suggests that within a reasonable sample size, the high-frequency components of noise stimuli mainly contribute to the  error of empirical cSTA. We speculate that the high-frequency components weakly perturb the spike timings, and thus, these components correlate non-significantly with stimuli eliciting a spike, $\xi\left(t_{s}^{i}-t\right)$. As the noise time constant increases, the high-frequency components of the noise stimuli become attenuated (Fig. \ref{fig4}), and the error becomes lower (Figs. \ref{fig1}(B1) and (B2)). These results are consistent with our view. On the other hand, the high-frequency components of the cSTA itself are also attenuated (Fig. \ref{fig4}). To estimate PRCs from empirical cSTAs, the algorithm needs to reconstruct the attenuated high-frequency components. However, the reconstruction algorithm simultaneously amplifies the error in the high-frequency components. Because of the trade-off between the error reduction of empirical cSTAs and the error amplification in retrieving PRCs, there is an optimal value of the noise time constant that obtain the best estimate of PRCs from empirical cSTAs (Figs. \ref{fig3} (B1) and (B2)).

One cannot a priori determine an optimal value of the noise time constant, $\tau$. However, our results suggest that by choosing a non-zero $\tau$, one can obtain a better solution than those by using white noise. Whereas one does not know the optimal $\tau$ a priori, one can obtain a semi-optimal solution by setting $\tau$ to be ten milliseconds. Furthermore, in many experiments with real biological cells, it is nearly impossible to stimulate with pure white noise, because measurement systems have a low-pass filter characteristic \cite{Brette}. By using our method, one can cancel out the low-pass filter effect of the measurement systems.

To theoretically evaluate the error in computing the PRC from the empirical cSTA, we need to analytically derive the second order statistics given by the spike-triggered covariance (STC) \cite{ota1,Arthur}. When deriving the STC analytically, we have to manipulate the four-body correlation of stimulus noise. In the case of white noise, it is possible to separate the four-body correlation into two-body correlations because white noise is uncorrelated, and thus, our research group succeeded in analytically deriving the STC. However, in the case of colored noise, it is difficult to perform such a separation. Therefore, at the present moment, we have no recipe for how to deal with the correlation case analytically.

The statistical algorithm proposed here can be straightforwardly extended to ad-hoc regression methods \cite{kota1} and Bayesian methods \cite{ota2}. We expect that these extended methods will ensure the accuracy of the estimate with a smaller sample size than that of the proposed method. However, the trade-off relation discussed above might remain unaltered by these extensions.

\section*{Acknowledgment}

This work was supported by MEXT KAKENHI Grant Number 26120512.

\newpage 
\bibliography{mori_cite1}

\begin{thebibliography}{22}
\expandafter\ifx\csname natexlab\endcsname\relax\def\natexlab#1{#1}\fi
\expandafter\ifx\csname bibnamefont\endcsname\relax
  \def\bibnamefont#1{#1}\fi
\expandafter\ifx\csname bibfnamefont\endcsname\relax
  \def\bibfnamefont#1{#1}\fi
\expandafter\ifx\csname citenamefont\endcsname\relax
  \def\citenamefont#1{#1}\fi
\expandafter\ifx\csname url\endcsname\relax
  \def\url#1{\texttt{#1}}\fi
\expandafter\ifx\csname urlprefix\endcsname\relax\def\urlprefix{URL }\fi
\providecommand{\bibinfo}[2]{#2}
\providecommand{\eprint}[2][]{\url{#2}}

\bibitem[{\citenamefont{Pikovsky et~al.}(2003)\citenamefont{Pikovsky,
  Rosenblum, and Kurths}}]{pikovsky}
\bibinfo{author}{\bibfnamefont{A.}~\bibnamefont{Pikovsky}},
  \bibinfo{author}{\bibfnamefont{M.}~\bibnamefont{Rosenblum}},
  \bibnamefont{and} \bibinfo{author}{\bibfnamefont{J.}~\bibnamefont{Kurths}},
  \emph{\bibinfo{title}{Synchronization: A Universal Concept in Nonlinear
  Sciences}} (\bibinfo{publisher}{Cambridge Univ.\ Press},
  \bibinfo{year}{2003}).

\bibitem[{\citenamefont{Buzsaki}(2006)}]{buzsaki}
\bibinfo{author}{\bibfnamefont{G.}~\bibnamefont{Buzsaki}},
  \emph{\bibinfo{title}{Rhythms of the brain}} (\bibinfo{publisher}{Oxford
  Univ.\ Press}, \bibinfo{year}{2006}).

\bibitem[{\citenamefont{Kuramoto}(1984)}]{kuramoto}
\bibinfo{author}{\bibfnamefont{Y.}~\bibnamefont{Kuramoto}},
  \emph{\bibinfo{title}{Chemical oscillations, waves and turbulance}}
  (\bibinfo{publisher}{Springer Verlag.}, \bibinfo{year}{1984}).

\bibitem[{\citenamefont{Ukai et~al.}(2007)\citenamefont{Ukai, Kobayashi,
  Nagano, Masumoto, Sujino, Kondo, Yagita, Shigeyoshi, and Ueda}}]{ukai}
\bibinfo{author}{\bibfnamefont{H.}~\bibnamefont{Ukai}},
  \bibinfo{author}{\bibfnamefont{T.~J.} \bibnamefont{Kobayashi}},
  \bibinfo{author}{\bibfnamefont{M.}~\bibnamefont{Nagano}},
  \bibinfo{author}{\bibfnamefont{K.~H.} \bibnamefont{Masumoto}},
  \bibinfo{author}{\bibfnamefont{M.}~\bibnamefont{Sujino}},
  \bibinfo{author}{\bibfnamefont{T.}~\bibnamefont{Kondo}},
  \bibinfo{author}{\bibfnamefont{K.}~\bibnamefont{Yagita}},
  \bibinfo{author}{\bibfnamefont{Y.}~\bibnamefont{Shigeyoshi}},
  \bibnamefont{and} \bibinfo{author}{\bibfnamefont{H.~R.} \bibnamefont{Ueda}},
  \bibinfo{journal}{Nat.\ Cell Biol.} \textbf{\bibinfo{volume}{9}},
  \bibinfo{pages}{1327} (\bibinfo{year}{2007}).

\bibitem[{\citenamefont{Reyes and Fetz}(1993)}]{reyes}
\bibinfo{author}{\bibfnamefont{A.~D.} \bibnamefont{Reyes}} \bibnamefont{and}
  \bibinfo{author}{\bibfnamefont{E.~E.} \bibnamefont{Fetz}},
  \bibinfo{journal}{J.\ Neaurophys.} \textbf{\bibinfo{volume}{69}},
  \bibinfo{pages}{1661} (\bibinfo{year}{1993}).

\bibitem[{\citenamefont{Schultheiss et~al.}(2012)\citenamefont{Schultheiss,
  Prinz, and Butera}}]{schultheiss}
\bibinfo{editor}{\bibfnamefont{N.~W.} \bibnamefont{Schultheiss}},
  \bibinfo{editor}{\bibfnamefont{A.~A.} \bibnamefont{Prinz}}, \bibnamefont{and}
  \bibinfo{editor}{\bibfnamefont{R.~J.} \bibnamefont{Butera}}, eds.,
  \emph{\bibinfo{title}{Phase response Curves in Neuroscience: Theory,
  Experiment, and Analysis}} (\bibinfo{publisher}{Springer},
  \bibinfo{year}{2012}).

\bibitem[{\citenamefont{Nielsen et~al.}(2010)\citenamefont{Nielsen, Uusisaari,
  and Stiefel}}]{nielsen}
\bibinfo{author}{\bibfnamefont{B.~T.} \bibnamefont{Nielsen}},
  \bibinfo{author}{\bibfnamefont{M.}~\bibnamefont{Uusisaari}},
  \bibnamefont{and} \bibinfo{author}{\bibfnamefont{M.}~\bibnamefont{Stiefel}},
  \bibinfo{journal}{Front.\ Neuroinform.} \textbf{\bibinfo{volume}{4}},
  \bibinfo{pages}{6.10.3389} (\bibinfo{year}{2010}).

\bibitem[{\citenamefont{Ermentrout et~al.}(2007)\citenamefont{Ermentrout,
  Galan, and Urban}}]{ermentrout1}
\bibinfo{author}{\bibfnamefont{G.~B.} \bibnamefont{Ermentrout}},
  \bibinfo{author}{\bibfnamefont{R.~F.} \bibnamefont{Galan}}, \bibnamefont{and}
  \bibinfo{author}{\bibfnamefont{N.~N.} \bibnamefont{Urban}},
  \bibinfo{journal}{Phys.\ Rev.\ Lett.} \textbf{\bibinfo{volume}{99}},
  \bibinfo{pages}{248103} (\bibinfo{year}{2007}).

\bibitem[{\citenamefont{Ota et~al.}(2009)\citenamefont{Ota, Nomura, and
  Aoyagi}}]{kota1}
\bibinfo{author}{\bibfnamefont{K.}~\bibnamefont{Ota}},
  \bibinfo{author}{\bibfnamefont{M.}~\bibnamefont{Nomura}}, \bibnamefont{and}
  \bibinfo{author}{\bibfnamefont{T.}~\bibnamefont{Aoyagi}},
  \bibinfo{journal}{Phys.\ Rev.\ lett.} \textbf{\bibinfo{volume}{103}},
  \bibinfo{pages}{024101} (\bibinfo{year}{2009}).

\bibitem[{\citenamefont{Dorval et~al.}(2001)\citenamefont{Dorval, Christini,
  and White}}]{dorval}
\bibinfo{author}{\bibfnamefont{A.~D.} \bibnamefont{Dorval}},
  \bibinfo{author}{\bibfnamefont{D.~J.} \bibnamefont{Christini}},
  \bibnamefont{and} \bibinfo{author}{\bibfnamefont{J.~A.} \bibnamefont{White}},
  \bibinfo{journal}{Annals of Biomedical Engineering}
  \textbf{\bibinfo{volume}{29}} (\bibinfo{year}{2001}).

\bibitem[{\citenamefont{Ota et~al.}(2011)\citenamefont{Ota, Omori, Watanabe,
  Miyakawa, Okada, and Aonishi}}]{ota2}
\bibinfo{author}{\bibfnamefont{K.}~\bibnamefont{Ota}},
  \bibinfo{author}{\bibfnamefont{T.}~\bibnamefont{Omori}},
  \bibinfo{author}{\bibfnamefont{S.}~\bibnamefont{Watanabe}},
  \bibinfo{author}{\bibfnamefont{H.}~\bibnamefont{Miyakawa}},
  \bibinfo{author}{\bibfnamefont{M.}~\bibnamefont{Okada}}, \bibnamefont{and}
  \bibinfo{author}{\bibfnamefont{T.}~\bibnamefont{Aonishi}},
  \bibinfo{journal}{Phys.\ Rev.\ E} \textbf{\bibinfo{volume}{84}},
  \bibinfo{pages}{041902} (\bibinfo{year}{2011}).

\bibitem[{\citenamefont{Nakao et~al.}(2010)\citenamefont{Nakao, Teramae,
  Goldobin, and Kuramoto}}]{nakao}
\bibinfo{author}{\bibfnamefont{H.}~\bibnamefont{Nakao}},
  \bibinfo{author}{\bibfnamefont{J.}~\bibnamefont{Teramae}},
  \bibinfo{author}{\bibfnamefont{D.~S.} \bibnamefont{Goldobin}},
  \bibnamefont{and} \bibinfo{author}{\bibfnamefont{Y.}~\bibnamefont{Kuramoto}},
  \bibinfo{journal}{Chaos} \textbf{\bibinfo{volume}{20}}
  (\bibinfo{year}{2010}).

\bibitem[{\citenamefont{Simakov and Perez-Mercader}(2013)}]{simakov}
\bibinfo{author}{\bibfnamefont{D.~S.~A.} \bibnamefont{Simakov}}
  \bibnamefont{and}
  \bibinfo{author}{\bibfnamefont{J.}~\bibnamefont{Perez-Mercader}},
  \bibinfo{journal}{J.\ Phys.\ Chemis.} \textbf{\bibinfo{volume}{117}}
  (\bibinfo{year}{2013}).

\bibitem[{\citenamefont{Kurebayashi et~al.}(2012)\citenamefont{Kurebayashi,
  Fujiwara, and Ikeguchi}}]{kurebayashi}
\bibinfo{author}{\bibfnamefont{W.}~\bibnamefont{Kurebayashi}},
  \bibinfo{author}{\bibfnamefont{K.}~\bibnamefont{Fujiwara}}, \bibnamefont{and}
  \bibinfo{author}{\bibfnamefont{T.}~\bibnamefont{Ikeguchi}},
  \bibinfo{journal}{EPL (Europhysics Letters)} \textbf{\bibinfo{volume}{97}},
  \bibinfo{pages}{50009} (\bibinfo{year}{2012}).

\bibitem[{\citenamefont{Schwartz et~al.}(2006)\citenamefont{Schwartz, Pillow,
  and Rust}}]{schwartz}
\bibinfo{author}{\bibfnamefont{O.}~\bibnamefont{Schwartz}},
  \bibinfo{author}{\bibfnamefont{J.~W.} \bibnamefont{Pillow}},
  \bibnamefont{and} \bibinfo{author}{\bibfnamefont{N.~C.} \bibnamefont{Rust}},
  \bibinfo{journal}{J.\ Vision} \textbf{\bibinfo{volume}{6}},
  \bibinfo{pages}{484} (\bibinfo{year}{2006}).

\bibitem[{\citenamefont{Ota et~al.}(2012)\citenamefont{Ota, T.Omori, Miyakawa,
  Okada, and Aonishi}}]{ota1}
\bibinfo{author}{\bibfnamefont{K.}~\bibnamefont{Ota}},
  \bibinfo{author}{\bibnamefont{T.Omori}},
  \bibinfo{author}{\bibfnamefont{H.}~\bibnamefont{Miyakawa}},
  \bibinfo{author}{\bibfnamefont{M.}~\bibnamefont{Okada}}, \bibnamefont{and}
  \bibinfo{author}{\bibfnamefont{T.}~\bibnamefont{Aonishi}},
  \bibinfo{journal}{Plos one} \textbf{\bibinfo{volume}{7}},
  \bibinfo{pages}{e50232} (\bibinfo{year}{2012}).

\bibitem[{\citenamefont{Risken}(1989)}]{risken}
\bibinfo{author}{\bibfnamefont{H.}~\bibnamefont{Risken}},
  \emph{\bibinfo{title}{The Fokker-Planck equation. Methods of solution and
  applications}} (\bibinfo{publisher}{Springer}, \bibinfo{year}{1989}).

\bibitem[{\citenamefont{Morris and Lecar}(1981)}]{morris}
\bibinfo{author}{\bibfnamefont{C.}~\bibnamefont{Morris}} \bibnamefont{and}
  \bibinfo{author}{\bibfnamefont{H.}~\bibnamefont{Lecar}},
  \bibinfo{journal}{Biophys.\ J.} \textbf{\bibinfo{volume}{35}},
  \bibinfo{pages}{193} (\bibinfo{year}{1981}).

\bibitem[{\citenamefont{Ermentrout}(1996)}]{ermentrout2}
\bibinfo{author}{\bibfnamefont{G.~B.} \bibnamefont{Ermentrout}},
  \bibinfo{journal}{Neural comput.} \textbf{\bibinfo{volume}{8}},
  \bibinfo{pages}{979} (\bibinfo{year}{1996}).

\bibitem[{\citenamefont{Bishop}(2006)}]{bishop}
\bibinfo{author}{\bibfnamefont{C.~H.} \bibnamefont{Bishop}},
  \emph{\bibinfo{title}{Pattern recognition and machine learning}}
  (\bibinfo{publisher}{Springer}, \bibinfo{year}{2006}).

\bibitem[{\citenamefont{Brette et~al.}(2008)\citenamefont{Brette, Piwkowska,
  Monier, Rudolph-Lilith, Fournier, Levy, Fregnac, Bal, and Destexhe}}]{Brette}
\bibinfo{author}{\bibfnamefont{R.}~\bibnamefont{Brette}},
  \bibinfo{author}{\bibfnamefont{Z.}~\bibnamefont{Piwkowska}},
  \bibinfo{author}{\bibfnamefont{C.}~\bibnamefont{Monier}},
  \bibinfo{author}{\bibfnamefont{M.}~\bibnamefont{Rudolph-Lilith}},
  \bibinfo{author}{\bibfnamefont{J.}~\bibnamefont{Fournier}},
  \bibinfo{author}{\bibfnamefont{M.}~\bibnamefont{Levy}},
  \bibinfo{author}{\bibfnamefont{Y.}~\bibnamefont{Fregnac}},
  \bibinfo{author}{\bibfnamefont{T.}~\bibnamefont{Bal}}, \bibnamefont{and}
  \bibinfo{author}{\bibfnamefont{A.}~\bibnamefont{Destexhe}},
  \bibinfo{journal}{Neuron} \textbf{\bibinfo{volume}{59}}, \bibinfo{pages}{379}
  (\bibinfo{year}{2008}).

\bibitem[{\citenamefont{Arthur et~al.}(2013)\citenamefont{Arthur, Burton, and
  Ermentrout}}]{Arthur}
\bibinfo{author}{\bibfnamefont{J.~G.} \bibnamefont{Arthur}},
  \bibinfo{author}{\bibfnamefont{S.~D.} \bibnamefont{Burton}},
  \bibnamefont{and} \bibinfo{author}{\bibfnamefont{G.~B.}
  \bibnamefont{Ermentrout}}, \bibinfo{journal}{J Comput Neurosci}
  \textbf{\bibinfo{volume}{34}}, \bibinfo{pages}{505} (\bibinfo{year}{2013}).

\end{thebibliography}

\end{document}